\documentclass[%
reprint,
superscriptaddress,
 amsmath,amssymb,
 aps,
 pra,
]{revtex4-2}

\usepackage{textgreek}
\usepackage{nicefrac}
\usepackage{siunitx}
\usepackage{multirow}
\usepackage[normalem]{ulem}
\usepackage{graphicx}
\usepackage{dcolumn}
\usepackage{bm}
\usepackage{xcolor}
\usepackage{siunitx}
\usepackage{changes}

\def\orcid#1{\kern .08em\href{https://orcid.org/#1}{\includegraphics[keepaspectratio,width=0.7em]{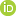}}}
\newcommand{\eg}{e.g.}

\newcommand{\ie}{i.e.}

\newcommand{\pstates}{$1s2p\,^3\!P_{0,1,2}$}

\newcommand{\lines}{$1s2s\,^3\!S_1\rightarrow\,1s2p\,^3\!P_{0,1,2}$}

\newcommand{\Cq}[1]{\ensuremath{^{#1}\text{C}^{4+}}}

\def\bdot{\mbox{\boldmath{$\cdot$}}}

\begin{document}


\title{Splitting Isotope Shift in the \pstates\ Fine-Structure Triplet in $^{12,13,14}$C$^{4+}$:\\
       Experiment and Theory}

\author{Patrick M\"{u}ller\orcid{0000-0002-4050-1366}}
  \affiliation{Institut für Kernphysik, Technische Universität Darmstadt, 64289 Darmstadt, Germany}

\author{Kristian König\orcid{0000-0001-9415-3208}}
  \email{kkoenig@ikp.tu-darmstadt.de}
  \affiliation{Institut für Kernphysik, Technische Universität Darmstadt, 64289 Darmstadt, Germany}%
  \affiliation{Helmholtz Research Academy Hesse for FAIR, GSI Helmholtzzentrum für Schwerionenforschung, 64291 Darmstadt, Germany}

\author{Emily Burbach\orcid{0009-0001-4237-0718}}
  \affiliation{Institut für Kernphysik, Technische Universität Darmstadt, 64289 Darmstadt, Germany}

\author{Gordon W. F. Drake\orcid{0000-0001-5116-1515}}
  \affiliation{Department of Physics, University of Windsor, Windsor, Ontario, Canada N9B 3P4}
    
\author{Phillip Imgram\orcid{0000-0002-3559-7092}}
  \affiliation{Institut für Kernphysik, Technische Universität Darmstadt, 64289 Darmstadt, Germany}
  
\author{Bernhard Maa\ss\orcid{0000-0002-6844-5706}}
  \altaffiliation{current address: Physics Division, Argonne National Laboratory, IL 60439 Lemont, USA}
  \affiliation{Institut für Kernphysik, Technische Universität Darmstadt, 64289 Darmstadt, Germany}

\author{Titamarie M. Maggio}
  \affiliation{Department of Physics, University of Windsor, Windsor, Ontario, Canada N9B 3P4}
    
\author{Wilfried N\"ortersh\"auser\orcid{0000-0001-7432-3687}}
\affiliation{Institut für Kernphysik, Technische Universität Darmstadt, 64289 Darmstadt, Germany}%
\affiliation{Helmholtz Research Academy Hesse for FAIR, GSI Helmholtzzentrum für Schwerionenforschung, 64291 Darmstadt, Germany}

\author{Julien Spahn\orcid{0009-0007-8354-4896}}
\affiliation{Institut für Kernphysik, Technische Universität Darmstadt, 64289 Darmstadt, Germany}%

\date{\today}

\begin{abstract}
We report measurements and theoretical calculations of the fine-structure splittings in all three \lines\ transitions in the heliumlike systems of the isotopes $^{12,13,14}$C. The metastable triplet state was efficiently populated in an electron beam ion source and the C$^{4+}$ ions were electrostatically accelerated to 50\,keV to perform collinear laser spectroscopy. From the determined transition frequencies, the splitting isotope shift (SIS), \ie, the difference in fine-structure splittings between different isotopes of the same element, was extracted. In the SIS, theoretical uncertainties due to higher-order quantum electrodynamic corrections are strongly suppressed since they are independent of both nuclear mass and the fine-structure quantum number $J$ in lowest order.
Comparison with theory provides an important test of experimental accuracy, particularly in the \Cq{13} case, for which the nuclear spin leads to hyperfine-induced fine-structure mixing. At the same time, the even-even isotopes \Cq{12,14} without nuclear spin can be used to confirm theory. Theoretical values of the SIS are given for all the heliumlike ions with $2\le Z\le 10$.        
\end{abstract}

\maketitle

\section{Introduction}
\label{sec:intro}
Heliumlike systems are the simplest atomic systems that include electron-electron correlation terms, which play an important role in high-precision calculations but are notoriously difficult to treat in \textit{ab initio} calculations \cite{Drake2020}. For systems at low $Z$, the relative contribution of the electron-electron interaction is large and calculations are performed starting from a solution of the nonrelativistic Schrödinger equation that is computed to very high numerical precision. Relativistic and quantum electrodynamic contributions are then calculated using a perturbation expansion in the parameters $\alpha$ and $Z\alpha$ with power-counting in the fine-structure constant $\alpha$ \cite{Haidar2020,Pachucki2010}. In this approach, called nonrelativistic quantum electrodynamics (NRQED), the electron-electron interaction is included to all orders. In high-$Z$ systems, on the contrary, the relativistic contributions grow with $Z\alpha$, while the relative contribution of the electron-electron interaction term diminishes. 
In this case, the starting point is solutions of the Dirac equation, including all powers of $Z\alpha$, while QED interactions are added in a series expansion in $\alpha$ and the electron-electron interaction is treated in a $1/Z$ power series \cite{Persson1996,Smits2023}. 

Both extremes, the He atom and high-$Z$ He-like systems have been investigated with increasing accuracy in the last decade, though at very different levels of relative accuracy. Examples are measurements of the transition frequencies and fine-structure splittings in stable helium isotopes \cite{Pastor.2004,Borbely.2009,Heydarizadmotlagh.2024} to test QED in light systems and to determine the fine-structure constant $\alpha$ \cite{Pachucki2010}, measurements of isotope shifts to extract changes in the nuclear charge radius of stable and short-lived isotopes \cite{Wang.2004,Mueller.2007,Lu.2013}, and measurements of transitions in heliumlike uranium to test QED in the strong-field regime \cite{Loetzsch.2024,Pfafflein.2025}.
These experiments have been accompanied or guided on the theoretical side, by increasingly complex and accurate calculations of the two-electron system in light \cite{Pachucki2010,Patkos.2021,Yerokhin.2018,Yerokhin.2022} to heavy ions \cite{Yerokhin.2022,Yerokhin.2015,Indelicato.2019,Malyshev.2023}.

In the intermediate region, in which both contributions become equally important, high-precision measurements and calculations have been scarce. Only a few measurements in He-like systems above Li have been reported previously, \eg, \cite{Myers.1999,Thompson.1998,Ozawa.2000}. However, with the advent of improved high-precision calculations for helium- and lithiumlike systems in the intermediate $Z$ region \cite{Yerokhin.2022,Puchalski.2013,Qi.2020} based on the ``unified method" \cite{Drake78,Drake1988} merging the low-$Z$ region with the high-$Z$ region, measurements in these systems have become of high interest and can be used to extract nuclear parameters, such as absolute values of the charge radius or the Zemach radius \cite{Imgram.2023,Qi.2020}. However, the new approaches have to be consolidated with respect to their accuracy and appropriate description of the physical systems. 

Recently, we have performed measurements of the transition frequencies of all three \lines\ fine-structure components in the heliumlike system of the most abundant even-even isotope \Cq{12} \cite{Imgram.2023} and found excellent agreement with the most advanced calculations for this system including all terms up to $m\alpha^7$ \cite{Yerokhin.2022}. 
In principle, these calculations enable the extraction of the nuclear charge radius solely from the measured transition frequency; however, the theoretical accuracy is still not sufficient to be  competitive with elastic electron scattering \cite{Cardman.1980, Sick.1982,Reuter.1982,Offermann.1991} or muonic atom X-ray spectroscopy \cite{Schaller.1982,Ruckstuhl.1984}. 
Comparing isotopes of the same element significantly reduces the uncertainty of NRQED calculations. Together with a measurement of the complete hyperfine structure in all fine-structure lines of \Cq{13}, the differential charge radius was improved by a factor three compared to muonic measurements \cite{Mueller.2025}.
Here, hyperfine-induced fine-structure mixing, a second-order hyperfine structure effect, had to be considered since it shifts individual hyperfine lines by up to \SI{2}{GHz}. 

We have now determined the fine-structure splittings in the radioactive isotope \Cq{14}. This allows us to compare the splitting isotope shift (SIS); \ie,  the difference in fine-structure splittings between different isotopes of the same element, with theoretical results \cite{Drake2005,Wang2017}. The SIS is determined by the relativistic finite nuclear mass and recoil contributions to the energy. Theoretical uncertainties due to higher-order quantum electrodynamic corrections are strongly suppressed since they are in lowest order independent of both nuclear mass and the fine-structure quantum number $J$. Therefore, agreement between experiment and theory provides an excellent and independent test for the experimental accuracy \cite{Nortershauser.2015} and the correct treatment of the hyperfine-induced fine-structure mixing.  For these reasons, the SIS has proven useful in detecting inconsistencies in experimental data, especially in cases where different measurements do not agree with each other as it was the case, \eg, in lithium \cite{Drake2005,Nortershauser.2011,Sansonetti.2011}. 

\section{Experiment}
\label{sec:Experiment}

\begin{figure*}[t]
    \centering
    \includegraphics[width=1\linewidth]{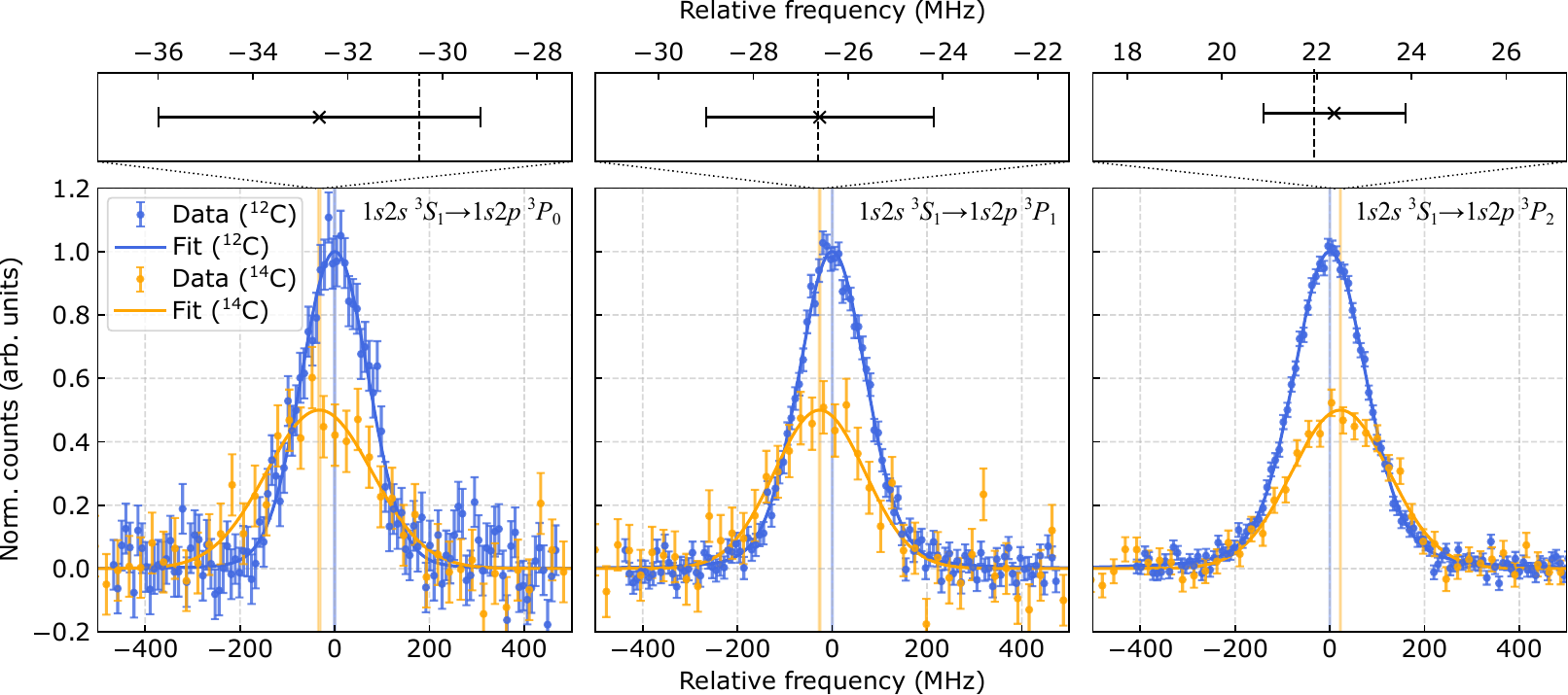}
    \caption{Typical spectra of the \lines\ transitions in \Cq{12,14}. The $y$-axis was normalized and the $x$-axis centered to the respective transition frequency of \Cq{12}. The resonance center of \Cq{14} is shifted by the SIS as defined in Eq.\,\eqref{Eq:CG_SIS}. The width of the vertical bars at the line centers represents the total uncertainty (statistical + systematic) of each transition frequency after averaging over all measurements. 
    At the top, a 10-MHz close-up depicts the experimentally and theoretically determined SIS. The total experimental uncertainty is depicted as error bar and the theoretical value (2\,kHz uncertainty) as vertical line.}
    \label{fig:Spectra}
\end{figure*}

All recent measurements in C$^{4+}$ were performed at the COllinear Apparatus for Laser Spectroscopy and Applied Sciences (COALA) \cite{Konig.2020}. The experimental procedure has been described in detail in \cite{ImgramPRA.2023} and only a short summary is given here.
The C$^{4+}$ ions were produced in an electron beam ion source (EBIS) using leaking mode operation; \ie, a continuous beam of ions was extracted. For the production of \Cq{12} natural propane gas was used. \Cq{13} was produced with enriched $^{13}$CH$_4$ and a 50:50 mixture of $^{12,14}$CO$_2$ was employed for the production of \Cq{14}. 
The EBIS trap potentials, electron current and gas pressure were optimized for the generation of the C$^{4+}$ charge state, which was 
selected according to its $q/m$ with a Wien filter.
Typical ion beams with currents of about 1.5\,nA of \Cq{12} and \Cq{13} and $0.8$\,nA of \Cq{14} were injected into the COALA beamline and superimposed with two cw laser beams---one in collinear and one in anticollinear geometry. 
Both laser beams, with wavelengths between 226.5 nm and 228.5\,nm, were generated using frequency-quadrupled Ti:sapphire lasers. The fundamental infrared laser frequencies at $\approx 910$\,nm were kept constant by an active stabilization to the beat signals with a frequency comb that was referenced to an atomic clock through a GPS-disciplined quartz oscillator.
In the central part of the beamline, two iris diaphragms in 2.6-m distance were used to check the overlap of the beams. In between the diaphragms the laser-ion-interaction region was located. It could be floated by a scan voltage (typically 10\,V) to change the ion velocity, which went along with a corresponding Doppler shift, and was equipped with a mirror- and a lens-based fluorescence-light detection system \cite{MuellerDiss.2024}. 
The laser frequencies in the laboratory frame $\nu_\mathrm{c/a}$ were chosen in such a way that the resonances appeared within $\Delta U=0.5$\,V of the scanning voltages for both geometries. 
Single photons were detected using photomultiplier tubes (PMT) to generate fluorescence spectra, such as shown in Fig.\,\ref{fig:Spectra} for \Cq{12,14}.
Collinear (c) and anticollinear (a) spectra were taken in quick succession.
Absolute transition frequencies $\nu_0$ were determined from the fitted resonant Doppler-shifted laser frequencies $\nu_\mathrm{c/a}$ using
\begin{align}
\nu_0 &= \sqrt{\nu_\mathrm{c}\nu_\mathrm{a}} - \frac{h\nu_\mathrm{c}\nu_\mathrm{a}}{2mc^2},
\label{eq:nu0}
\end{align}
where the subtracted term accounts for the transferred recoil of the absorbed photon, with the ion mass $m$, the Planck constant $h$ and the vacuum speed of light $c$.

The main uncertainty of the COALA experiment originates from the horizontal variation in the overlap of the collinear with the anticollinear laser beam and the asymmetric velocity distribution in the ion-beam cross-section due to the electrostatic deflection necessary to align ion and laser beams.
Any imperfect horizontal laser overlap thus induces a shift of the measured scanning voltage difference $\Delta U$, due to the interaction with different sets of ion velocities. Since the laser overlap was readjusted regularly, this effect appears as a symmetric statistical uncertainty and was estimated based on systematic measurements and simulations to be 1.7\,MHz \cite{ImgramPRA.2023}. At least 50 collinear-anticollinear measurements were taken for each of the three transitions at three or more days with independent laser adjustments.
The statistical uncertainty is determined as the standard error of the mean and is $\approx 1$\,MHz for all transitions.
Additional uncertainties originate from multiple photon momentum transfers before the considered final absorption, Zeeman shifts, and an angular misalignment of the ion and laser beams and are below 0.4\,MHz. For detailed discussions of the experimental uncertainties, see \cite{ImgramPRA.2023, Mueller.2025}. All contributions are added in quadrature and the total uncertainty is plotted as the width of the vertical bars at the line centers in Fig.\,\ref{fig:Spectra}. The center frequencies of \Cq{14} are shifted relative to those of \Cq{12} in accordance with the determined SIS.\\
The SIS was determined from the measured rest-frame frequencies $\nu_{P0}^A, \nu_{P1}^A, \nu_{P2}^A$ of the three \lines\ fine structure transitions in the isotopes \Cq{12,13,14} with $A=12,13,14$ as
\begin{equation}
\label{Eq:CG_SIS}
    \mathrm{SIS}_{P_J}^A=\nu_{P_J}^A-\nu_{\rm cg}^A- (\nu_{P_J}^{12}-\nu_{\rm cg}^{12}).
\end{equation}
with the center-of-gravity frequency
\begin{equation}
    \nu_{\rm cg}^A = \frac{1}{9} (\nu_{P_0}^A+3\nu_{P_1}^A+5\nu_{P_2}^A).
\end{equation}
and \Cq{12} as the reference isotope. The upper part of Fig.\,\ref{fig:Spectra} shows the experimentally extracted SIS along with its total uncertainty, which includes the contributions from both $^{12}$C and $^{14}$C. The dashed line represents the theoretical prediction, with an associated uncertainty of 2\,kHz.



\section{Splitting isotope shift theory}
\label{sec:Theory}
As noted in the Introduction, the splitting isotope shift (SIS) \cite{Yan.2008} refers to the isotope shift in the fine-structure splitting for the $1s2p\;^3\!P_J$ states.
It provides a strong internal consistency check of the experimental data, and a test of hyperfine averaging in the case of $^{13}$C.
A particular advantage of the SIS is that it can be accurately calculated from the lowest-order relativistic recoil part of the Breit interaction, without significant uncertainties from nuclear size contributions or higher-order QED effects.

The theoretical formulation begins with an expansion of the total energy $E$ in powers of the fine structure constant $\alpha \simeq 1/137.0359991$ and the ratio $\mu/M$ of the reduced electron mass $\mu=M m_{\rm e}/(M+m_{\rm e})$ and nuclear mass $M$.
The atomic energies can then be expanded in the form
\begin{equation}
E = E_{\rm nr}  + \alpha^2E_{\rm rel} + \alpha^3E_{\rm QED} + \cdots
\end{equation}
within NR-QED, where each $E_X$ has the expansion
\begin{equation}
E_X = E_X^{(0)} + \frac{\mu}{M}E_X^{(1)} + \left(\frac{\mu}{M}\right)^2E_X^{(2)} + \cdots
\end{equation}
The key point of the analysis is that all the leading $E_X^{(0)}$ terms cancel from the SIS.
As a consequence, higher order theoretical uncertainties are suppressed by a factor of
$\frac{\mu}{M_{A}} - \frac{\mu}{M_{A'}}$, which is $3.5\times10^{-6}$ for the $^{13}$C -- $^{12}$C isotope shift and $6.5\times10^{-6}$ for the $^{14}$C -- $^{12}$C isotope shift.
%

The starting point for a theoretical formulation of the SIS is the nonrelativistic Hamiltonian for a heliumlike ion, obtained by transforming from an inertial frame to center-of-mass plus relative coordinates.   In order to keep track of the mass scaling of various quantities, it is convenient to work in reduced-mass atomic units (a.u.) with $\hbar=\mu=e=1$. The unit of distance is the reduced-mass Bohr radius $a_\mu = (\mu/m_{\rm e})a_0$, and the unit of energy is the reduced mass Hartree  $\SI{1}{E_{\rm h}}=e^2/a_\mu$. 
In the absence of external forces, the Hamiltonian for relative motion for a heliumlike ion with nuclear charge $Ze$ is then (with $4\pi\epsilon_0=1$)
\begin{equation}
H = \frac{p_1^2}{2} + \frac{p_2^2}{2} + \left(\frac{\mu}{M}\right){\bf p}_1{\bf\cdot p}_2 - \frac{Z}{r_1} - \frac{Z}{r_2} + \frac{1}{r_{12}} .
\end{equation}
where ${\bf r}_1$ and ${\bf r}_2$ are the position vectors of the two electrons relative to the nucleus, and $r_{12} = |{\bf r}_1 - {\bf r}_2|$ is the inter-electron separation.  This form of the
Hamiltonian clearly reveals the role of the mass polarization operator $H_{\rm MP}= \left(\frac{\mu}{M}\right){\bf p}_1{\bf\cdot p}_2$ as a perturbation relative to the infinite nuclear mass case, with $\mu/M$ serving as the perturbation parameter.  The final results are then converted to physical units (MHz) with use of the reduced mass Hartree E$_{\rm h}$ separately for each isotope before calculating the SIS.

After cancelling the mass-independent terms, the remaining nonvanishing contributions to the SIS can be written in the form
\begin{equation}
\Delta E(\mbox{SIS}) = \Delta E_{\rm nr} + \Delta E_{\rm rel} + \Delta E_{\rm QED}
\end{equation}
where the nonrelativistic energy has the $\mu/M$ expansion
\begin{equation}
 \Delta E_{\rm nr} = \left(\frac{\mu}{M}\right) E_{\rm nr}^{(1)} + \left(\frac{\mu}{M}\right)^2E_{\rm nr}^{(2)} +\mathcal{O}\left(\frac{\mu}{M}\right)^3
\end{equation}
and the relativistic recoil and radiative recoil terms have the expansions (to sufficient accuracy)
\begin{equation}
\Delta E_{\rm rel} = \left(\frac{\mu}{M}\right)E_{\rm rel}^{(1)} + \left(\frac{m_{\rm e}}{M}\right)E_{\rm Stone}+ \mathcal{O}\left(\frac{\mu}{M}\right)^2
\end{equation}
where $E_{\rm Stone}$ is a relativistic recoil term \cite{Stone1963} discussed further below, and 
\begin{equation}
\Delta E_{\rm QED} = \left(\frac{\mu}{M}\right)E_{\rm QED}^{(1)} + \mathcal{O}\left(\frac{\mu}{M}\right)^2.
\end{equation}

For example,  with $Z = 6$ and $n = 2$, the next higher-order QED term is of order
\begin{eqnarray*}
\alpha^4\left(\frac{\mu}{M_{12}} - \frac{\mu}{M_{13}}\right)\times\frac{(Z-1)^6}{n^3} &\simeq& 1.93\times10^{-11}\,\mbox{E$_{\rm h}$}\\
 &\simeq& 0.127\,\mbox{MHz}
\end{eqnarray*}
which is an order of magnitude less than the experimental uncertainty. We greatly reduce the uncertainty from this term by calculating the dominant part due to singlet-triplet mixing.  The numerical value of the remainder is then very small, and is the dominant source of uncertainty.

There is a further cancellation since, by definition, the SIS depends only on the spin-dependent parts of the above terms.  Therefore, $\Delta E_{\rm nr}$ does not contribute, leaving the leading terms $(\mu/M)E_{\rm rel}^{(1)}$ and $(\mu/M)E_{\rm QED}^{(1)}$ of $\Delta E_{\rm rel}$ and $\Delta E_{\rm QED}$ respectively as the only significant contributions to the SIS.  In addition, $\Delta E_{\rm QED}$ contributes only through the electron anomalous magnetic moment since the other lowest-order terms (in $\alpha$) are spin-independent.

The detailed expressions for the spin-dependent contributions to relativistic recoil are as follows \cite{Bethe}:  First, the
 $\Delta E_{\rm rel}$ terms of order $\alpha^2\mu/M$ E$_{\rm h}$ are
\begin{equation}
\label{Erel}
\left(\frac{\mu}{M}\right)E_{\rm rel}^{(1)} = \left(\frac{\mu}{M}\right)(E_{\rm scale}  + E_{\rm MP} + E_{\rm s-t}) + \left(\frac{m_{\rm e}}{M}\right) E_{\rm Stone}
\end{equation}
where the mass scaling of the usual Breit interaction from the expansion of $(1-\mu/M)^3$ is
\begin{equation}
E_{\rm scale} = -3\left(\langle B_{\rm so}\rangle + \langle B_{\rm soo}\rangle + \langle B_{\rm ss}\rangle\right)
\end{equation}
where the spin-orbit, spin-other-orbit and spin-spin terms are given by
\begin{eqnarray}
\label{Bso}
&B_{\rm so} = \frac{Z\mu_{\rm B}e(1+2a_{\rm e})}{mc}\left[\frac{{\bf r}_1\times{\bf p}_1\bdot{\bf s}_1}{r_1^3} + \frac{{\bf r}_2\times{\bf p}_2\bdot{\bf s}_2}{r_2^3}\right]\\
\label{Bsoo}
&B_{\rm soo} = -\frac{\mu_{\rm B}e}{2mcr_{12}^3}{\bf r}_{12}\times\left[(3+4a_{\rm e}){\bf p}_-\bdot{\bf s}_+ -
{\bf p}_+\bdot{\bf s}_-\right]\\
\label{Bss}
&B_{\rm ss} = 4\mu_e^2\left[\frac{8\pi}{3}\delta(r_{12}){\bf s}_1\bdot{\bf s}_2+
\frac{{\bf s}_1\bdot{\bf s}_2}{r_{12}^3} - \frac{3({\bf s}_1\bdot{\bf r}_{12}){\bf s}_2\bdot{\bf r}_{12}}{r_{12}^5}\right],
\end{eqnarray}
with ${\bf p}_\pm = {\bf p}_1 \pm {\bf p}_2$ and ${\bf s}_\pm = {\bf s}_1 \pm {\bf s}_2$. The electron anomalous magnetic moment is accounted for in the above through the $a_{\rm e} \simeq \alpha/(2\pi)$ term.  
The Stone term \cite{Stone1963} is defined by $E_{\rm Stone} = \langle \tilde\Delta_{\rm so}\rangle$ with
\begin{equation}
\tilde\Delta_{\rm so}=\frac{2Z\mu_{\rm B}e}{mc}\sum_{i=1} ^2 \frac{1}{r_i^3}{\bf{r}}
_i \times {\bf{p_+}} \bdot {\bf{s}}_i\,.
\label{eq:010}
\end{equation}
It arises from a transformation of the spin-orbit interaction to center-of-mass plus relative coordinates.  It contributes to the leading term of the SIS in lowest order $\alpha^2m_{\rm e}/M$. 
Next, the mass polarization operator $H_{\rm MP}$ introduces wave function corrections.  The corresponding energy shift
could in principle be calculated as second-order cross terms of the form
\begin{equation}
\frac{\mu}{M}\Delta E_X = 2\langle H_{\rm MP} (H_0 - E_0)^{\prime-1} B_X\rangle
\end{equation}
where $(H_0-E_0)^{\prime-1}$ is the resolvent operator, and $X$ refers to each of the spin-dependent terms in the Breit interaction (\ref{Bso}-\ref{Bss}), including the Stone term.
Instead, as a matter of expedience, we estimate both these terms, and the next-to-leading terms of order $\alpha^2(\mu/M)^2$ by writing them in the form
\begin{equation}
\label{quadratic}
\frac{\mu}{M}\Delta E_X = \frac{\mu}{M}\left( \Delta E_X^{(1)} + \frac{\mu}{M}\Delta E_X^{(2)}\right)
\end{equation}
and determining $\Delta E_X^{(1)}$ and $\Delta E_X^{(2)}$ by an exact quadratic fit to a pair of calculations obtained by repeating the entire calculation with the $H_{\rm MP}$ term included explicitly in the Hamiltonian for the case of $^{12}$C relative to the infinite mass case, and a second case with a value of $\mu/M$ arbitrarily chosen to be four times larger. This then gives a pair of equations that can be solved for the two parameters $\Delta E_X^{(1)}$ and $\Delta E_X^{(2)}$.  The next term in the series of order $\alpha^2(\mu/M)^3$ is too small to affect the calculation.  However, the quality of the convergence must be extremely good in order for the method to give meaningful results.

The final contribution to the SIS comes from singlet-triplet mixing due to the spin-dependent terms in the Breit interaction taken to second order. Since $J$ is a good quantum number (in the absence of hyperfine mixing), it is only the $2\;^3\!P_1$ state that mixes with the $^1\!P_1$ states. The general form of the correction for the $2\;^3\!P_1$ state, expressed as a spectral sum over states, is
\begin{equation}
\label{eq:pert}
\Delta E_{\rm rel}^{(2)}(2\;^3\!P_1) = \sum_{n=2}^\infty\frac{\langle 2\;^3\!P_1| B| n\;^1\!P_1\rangle\langle n\;^1\!P_1 | B |2\;^3\!P_1\rangle}{E(2\;^3\!P_1) - E(n\;^1\!P_1)}
\end{equation}
including an integration over the $^1\!P_1$ continuum, and similarly for the $2\;^1\!P_1$ state.  Although all intermediate states contribute, the sum is dominated by just the $n = 2$ term since the energy denominator $E(2\;^3\!P_1) - E(n\;^1\!P_1)$ is particularly small for this case, and it increases only in proportion to $Z$, instead of the usual $Z^2$ for all the other terms.  The order of magnitude is thus $\alpha^4Z^7$ E$_{\rm h}$ for the $n=2$ term, instead of $\alpha^4 Z^6$ E$_{\rm h}$ for all the others.  Our strategy is to subtract out the $n=2$ contribution from the second-order perturbation sum and replace it by an exact diagonalization within the $n = 2$ manifold of states.  This strategy avoids saturation of the singlet-triplet mixing with increasing $Z$, and accommodates the progressive transition from $LS$- to $jj$-coupling \cite{Drake78}. It is an implicit part of the unified method \cite{Drake78,Drake1988}.  The small contribution from the background of all 
other intermediate states is then added back in as a first-order perturbation, together with the spin-dependent Douglas and 
Kroll terms \cite{DouglasKroll,Drake2002}, as further discussed in Sec.\,\ref{results}. The linear and quadratic mass polarization contributions to the SIS from $E_{\rm s-t}$ are then calculated by a quadratic fit to a pair of isotopes, as described in Sec.\,\ref{sec:Theory} (see Eq.\,\eqref{quadratic}).

The spin-independent QED corrections do not contribute to the SIS, but for the sake of completeness, the formalism is included here.  
As discussed in detail in previous papers \cite{Noertershaeuser.2011,Lu.2013}, the
lowest order QED contribution to the energy, including radiative recoil, can be
expressed in the form
\begin{equation}
E_{\rm QED} = E_{\rm L,1} + E_{\rm M,1} + E_{\rm R,1} + E_{\rm L,2}
\end{equation}
where $E_{\rm L,1}$ is the mass-independent part of the
electron-nucleus Lamb shift (the Kabir-Salpeter term \cite{Kabir1957}),
$E_{\rm M,1}$ contains mass scaling and mass polarization corrections,
$E_{\rm R,1}$ contains recoil corrections (including radiative recoil),
and $E_{\rm L,2}$ is the electron-electron term originally obtained by
Araki \cite{Araki1957} and Sucher \cite{Sucher1958}. We are concerned
here primarily with the mass-dependent terms $E_{\rm M,1}$ and $E_{\rm R,1}$.  With
the notation $\langle\sum_i \delta({\bf r}_i)\rangle = \langle\sum_i \delta({\bf r}
_i)\rangle^{(0)} + \lambda\langle\sum_i \delta({\bf r}_i)\rangle^{(1)}+
\cdots$ to express the mass dependence of the $\delta$-function matrix
element, they are given by
\begin{eqnarray}
\label{E_L}
E_{\rm L,1} &=& \frac{4Z\alpha^3\langle\textstyle{\sum_i}\delta({\bf r}_i)\rangle^{(0)}}
{3}\left\{\ln(Z\alpha)^{-2} - \beta(2\;^3\!P) + \frac{19}{30}\right\}\nonumber\\ 
&&\mbox{}+ \mathcal{O}(\alpha^4)
\end{eqnarray}
\begin{eqnarray}
\label{E_M}
E_{\rm M,1} &=& \frac{\mu\langle\textstyle{\sum_i}\delta({\bf r}_i)\rangle^{(1)}}{M
\langle\textstyle{\sum_i}\delta({\bf r}_i)\rangle^{(0)}}E_{\rm L,1}\nonumber\\
&&\mbox{} + \frac{4Z\alpha^3\mu\langle\textstyle{\sum_i}\delta({\bf r}_i)\rangle^{(0)}}
{3M}\left[1 - \Delta\beta_{\rm MP}(2\;^3\!P)\right]~~~~~~~~
\end{eqnarray}
and
\begin{eqnarray}
E_{\rm R,1} &=& \frac{4Z^2\alpha^3\mu\langle\textstyle{\sum_i}\delta({\bf r}_i)
\rangle^{(0)}} {3M}\left[ \frac{1}{4}\ln(Z\alpha)^{-2} - 2\beta(2\;^3\!P) \right.\nonumber\\
&&\mbox{}\left. - \frac{1}{12} - \frac{7}{4}a(1s^2\,nL) \right] \,.
\label{E_R}
\end{eqnarray}
where $\beta(2\;^3\!P)$ is the Bethe logarithm for the $1s2p\;^3\!P$ state of C$^{4+}$, and
$\Delta\beta(2\;^3\!P)_{\rm MP}$ is the mass polarization correction.  The numerical values for
these quantities are  $\beta(2\;^3\!P) = 2.981\,835\,92(3)$ and $\Delta\beta(2\;^3\!P)_{\rm MP}
= 0.02398(1)$ as calculated in Ref.\ \cite{Drake_Goldman99}.
The connection with the corresponding terms in the hydrogenic
Lamb shift \cite{Eides2001} for the above terms can be seen by observing that an overall multiplying factor
of $\langle\delta({\bf r})\rangle = Z^3\delta_{L,0}/(\pi n^3)$ for the
hydrogenic case is here replaced by the correct expectation value
$\langle\sum_{j=1}^N\delta({\bf r}_j)\rangle$ for the multi-electron
case, summed over the $N=2$ electrons. The term $a(1s\,nL)$ corresponds to
a well-known term in the hydrogenic Lamb shift.  Its two-electron
generalization is \cite{YanDrake2002,Pachucki1998, Pachucki2000}
\begin{equation}
a(1s\,nL) = \frac{2Q_1^{(0)}}{\langle\textstyle{\sum_i}\delta({\bf r}_i)\rangle^{(0)}} + 2\ln Z - 3
\end{equation}
where
\begin{equation}
\label{Q1}
Q_1^{(0)} = \frac{1}{4\pi}\lim_{\epsilon\rightarrow 0}\sum_{i} \langle
r_{i}^{-3}(\epsilon)+ 4 \pi (\gamma_{\rm eu} +\ln \epsilon) \delta({\bf r}
_{i})\rangle\,
\end{equation}
and $\gamma_{\rm eu}$ is Euler's constant.
The residual state dependence due
to other terms such as the Bethe logarithm is then
relatively weak. 

The electron-electron QED shift $E_{\rm L,2}$ can similarly be
separated into mass-independent and mass dependent parts according to
\begin{equation}
E_{\rm L,2} = E_{\rm L,2}^{(0)} + \frac{\mu}{M}E_{\rm L,2}^{(1)} +
\cdots
\end{equation}
where 
\begin{eqnarray}
E_{\rm L,2}^{(0)} &=& \alpha^3\left(\frac{14}{3}\ln\alpha +\frac{164}{15}
- \pi\alpha\ln\alpha \right) \langle\delta({\bf r}_{12})
\rangle^{(0)} \nonumber\\
&&\mbox{}- \frac{14} {3}\alpha^3 Q^{(0)}
\end{eqnarray}
and the mass scaling and mass polarization corrections are
\begin{eqnarray}
\label{EL21}
E_{\rm L,2}^{(1)} &=& \alpha^3\left(\frac{14}{3}
\ln\alpha + \frac{164}{15}\right) \langle\delta({\bf r}_{12})
\rangle^{(1)} \nonumber\\
&&\mbox{}-\frac{14}{3}\alpha^3 \left(Q^{(1)} +\langle\delta({
\bf r}_{12})\rangle^{(0)}\right)
\end{eqnarray}
Following our notation, the $Q^{(0)}$-term for infinite mass is given
by
\begin{equation}
\label{Q0}
Q^{(0)} = \frac{1}{4\pi}\lim_{\epsilon\rightarrow 0} \langle
r_{12}^{-3}(\epsilon)+ 4 \pi (\gamma_{\rm eu} +\ln \epsilon) \delta({\bf r}
_{12})\rangle\,
\end{equation}
The $Q^{(1)}$ term is the correction due to the mass polarization
correction to the wave function and mass scaling. Note that the
infinitesimal limiting quantity $\epsilon$ has dimensions of distance,
and so it generates the additional term $
\langle\delta({\bf r}_{12})\rangle^{(0)}$ in Eq.\,\eqref{EL21}
when distances
are rescaled for the finite mass case according to $\epsilon
\rightarrow (\mu/m)\epsilon$.  The matrix element $\langle\delta({\bf r}_{12})\rangle$
vanishes for triplet states.    

\section{Calculations}
The main computational step is the evaluation of matrix elements of the operators as identified in the previous section, using suitable wave functions for the $P$-states of two-electron C$^{4+}$.  As in previous works, the heliumlike Hamiltonian is diagonalized in a basis set constructed from basis functions of the form (for $P$-states) \cite{DrakeYan92}
 \begin{eqnarray}
 \phi_{i,j,k}(\alpha,\beta;{\bf r}_1,{\bf r}_2) &=& r_1^i\,r_2^{j+1}\,r_{12}^k\exp(-\alpha r_1 - \beta r_2)\cos\theta_2\nonumber\\
                                                &&\mbox{}\pm \mbox{ exchange}
 \end{eqnarray}
where $\cos\theta_2$ gives the angular dependence on ${\bf r}_2$ appropriate to a $p$-electron.  The ``exchange" term denotes an interchange of the labels 1 and 2, with the (+) sign for triplets and the ($-$) sign for singlets.  A variational wave function in a double  basis set then has the form
\begin{equation}
\psi({\bf r}_1,{\bf r}_2) = \sum_{p=1}^{2}\,\,\sum_{i,j,k}^{i+j+k \le \Omega} c_{i,j,k}^{(p)}\phi_{i,j,k}(\alpha_p,\beta_p;{\bf r}_1,{\bf r}_2)
\end{equation}
where the $c_{i,j,k}^{(p)}$ are linear variational coefficients, and $\Omega$ controls the size of the basis set such that $i+j+k\le \Omega$.   The four nonlinear variational parameters $\alpha_p$, $\beta_p$ are determined by minimizing the energy on a four-dimensional energy surface.  This is accomplished by calculating analytically the derivatives $\partial E/\partial \alpha_p$, $\partial E/\partial \beta_p$ \cite{DrakeMakowski88} and finding their zeros by Newton's method. The optimization produces a natural separation of the nonlinear parameters into short-range, and asymptotic long-range sectors The basis set also includes the screened hydrogenic term
$\psi_{1s}({\bf r}_1;Z)\psi_{2p}({\bf r}_2;Z-1) \pm$ exchange for effective nuclear charges $Z$ and $Z-1$ respectively.

\begin{table*}[htb]
\caption{Contributions to the $1s2p\;^3\!P_J$ $^{13}$C$^{4+}-^{12}$C$^{4+}$ isotope shift in the (negative) binding energy (MHz).}
\label{tableI}
\begin{tabular}{l r@{}l r@{}l r@{}l}
\hline
\hline
\rule{0mm}{3mm}Term  &\multicolumn{2}{c}{$2\;^3\!P_0$}&\multicolumn{2}{c}{$2\;^3\!P_1$}&  \multicolumn{2}{c}{$2\;^3\!P_2$}\\
\hline
n.r.\ mass scaling         &$ -74806$&.64536(8) &$ -74806$&.64536(8) &$ -74806$&.64536(8)  \\  
1st.\ order mass pol.      &$  45950$&.33669(5) &$  45950$&.33669(5) &$  45950$&.33669(5)  \\  
2nd.\ order mass pol.      &$      8$&.241876(1)&$      8$&.241876(1)&$      8$&.241876(1) \\  
$\alpha^2\mu/M$ relativistic recoil  &$   -112$&.3674   &$   -106$&.1988  &$    -43$&.7862    \\  
 $\alpha^2\mu/M$ rel.\ mass pol.\    &$     46$&.0207(1)  &$     44$&.6923(1)  &$      5$&.7639(1)   \\  

$\alpha^3\mu/M$ anom.\ mag.\ moment        &$     -0$&.0206   &$     -0$&.0192   &$      0$&.0157    \\  
$\alpha^3\mu/M$ $E_{\rm L,1} + E_{\rm M,1} + E_{\rm R,1}$   &$      2$&.7492   &$      2$&.7492   &$      2$&.7492    \\  
$\alpha^3\mu/M$ $E_{\rm L,2}$  &$     -0$&.0131   &$     -0$&.0131   &$     -0$&.0131    \\  
$\alpha^2(\mu/M)^2$ second-order M.P.&$-0$&.0074   &$     -0$&.0071   &$       -0$&.0116   \\ 
$\alpha^2(\mu/M)^2$ mass scaling &$-0$&.0071   &$     -0$&.0023   &$       0$&.0028   \\ 
$\alpha^4\mu/M + \cdots$ S.I.\footnote{S.I. refers to the (approximately) spin-independent remainder due to higher-order terms in a hydrogenic approximation.}  &$-0$&.0374&$    -0$&.0376&$     -0$&.0380  \\  
$\alpha^4\mu/M$ $n=2$ sing.-trip.\ mixing  &$      0$& .0000  &$     -2$&.7676   &$      0$&.0000    \\  
$\alpha^4\mu/M$ remainder\footnote{Relative to $J=0$ \cite{Pachucki2010}.}  &$      0$&.0000   &$     -$0&.021(1)      &$       -0$&.001(1)    \\  
Finite nuclear size\footnote{The nuclear charge radii used as input parameters for the calculations are 2.4717(42) fm and 2.4464(45) fm for $^{12}$C and $^{13}$C, respectively \cite{Mueller.2025}.  The large uncertainty of $\pm$0.7\,MHz cancels when calculating the SIS.}        &$       2$&.8(7)    &$      2$&.8(7)  &$      2$&.8(7)      \\  
Total                      &$ -28908$&.966(1)   &$ -28906$&.910(1)   &$ -28880$&.601(1)    \\  
Centroid                   &         &          &$ -28892$&$.5(7)$ \\
\hline\hline
\end{tabular}
\end{table*}

The rationale for improved accuracy with a double basis set is that a point of diminishing returns (and numerical instability) is reached with increasing $\Omega$, and so doubling the basis set allows more terms to be added without $\Omega$ becoming excessively large.

The principal computational step  to obtain nonrelativistic wave functions and energies is to solve the generalized eigenvalue problem
\begin{equation}
{\bf H}\Psi = \lambda{\bf O}\Psi
\end{equation}
where {\bf H} is the Hamiltonian matrix with matrix elements $\langle\phi_{i^\prime,j^\prime,k^\prime}|H|\phi_{i,j,k}\rangle$,  {\bf O} is the overlap matrix in the nonorthogonal basis set, and $\Psi$ is the column vector of basis function coefficients $c_{i,j,k}^{(p)}$.   The power method is used to find the eigenvalue closest to an initial guess.  The strategy is to perform a sequence of calculations with increasing values of $\Omega$ and assess the rate of convergence to determine the accuracy.  As a matter of practical convenience, the power-method is used to find the eigenvalue closest to an initial guess in place of total diagonalization.

The actual basis sets used in the calculations have two important modifications relative to a simple doubling.  
First, for the special case of $P$-states, it is advantageous to include a third sector with the roles of the $s$- and $p$-electrons interchanged.  
This takes into account mixing of these terms by the ${\bf p}_1\bdot{\bf p}_2/M$ mass polarization interaction. Second, since each combination of powers $i,j,k$ is included three times in all, judicious truncations can be introduced to reduce the total size of the basis set. 
The truncations are: (1) for a given overall $\Omega$, the values in the three sectors are $\Omega_1 = \Omega$, $\Omega_2 = 2$, and $\Omega_3 =\Omega$; (2) in sectors 1 and 2, terms are omitted with high powers of $r_1$ and $r_{12}$ such that $i > 10$ and $j > 8$; and (3) in sector 3, terms are omitted for which $i+j+k + |i-j| \ge \Omega_i$ for $k > 2$ \cite{Kono83}. These 
truncations do not affect the ultimate convergence since all powers are eventually included as $\Omega$ increases.  However, for the largest $\Omega = 16$, it reduces the nominal basis set size of 2907 terms to a more modest 1521 terms. The converged nonrelativistic eigenvalue for the $1s2p\;^3\!P$ state of C$^{4+}$ for the case of infinite nuclear mass is  
\[
 -21.221\,710\,696\,488\,051\,044(19)
\]
These basis sets have excellent numerical stability so that standard quadruple precision (32 decimal digit) is sufficient.

\section{Results}
\label{results}
Table \ref{tableI} lists the theoretical contributions to the total isotope shifts in the binding energy for the $1s2p\;^3\!P_J$ states of $^{13}$C$^{4+}$ relative to $^{12}$C$^{4+}$ with $J=0,1,2$. It is included both to show the large extent of numerical cancellation that takes place in calculating the SIS, and to check that the isotope shift in the $2\,^3\!S_1-2\,^3\!P_{\rm cg}$ transition frequency agrees with experiment. The corresponding isotope shift for the binding energy of the $2\,^3\!S_1$ state is $-80638.1$\,MHz, resulting in a frequency shift of 51745.6\,MHz for the $2\,^3\!S_1-2\,^3\!P_{\rm cg}$ transition, in excellent agreement with the experimental value 51746.1(1.4)\,MHz \cite{Mueller.2025,MuellerDiss.2024}. 

\begin{table}[b]
\caption{Contribution of $2\,^1\!P_1-2\,^3\!P_1$ singlet-triplet mixing corrections to the isotope shift calculated by exact diagonalization ($E_{\rm{s-t},n=2}^{\rm exact}$) and by perturbation theory ($E_{\rm{s-t},n=2}^{\rm pert.}$) for heliumlike ions with nuclear charge $Z$, in units of $\alpha^4Z^8\mu/M$ a.u.  The last column is the difference $E_{\rm{Total}}^{\rm pert.}-E_{\rm{s-t},n=2}^{\rm pert.}$, with $E_{\rm{Total}}^{\rm pert.}$ from Ref.\ \cite{Pachucki2010}.}
\label{tableII}
\begin{tabular}{r r@{}l r@{}l r@{}l r@{}l }
\hline
\hline
$Z$ &  \multicolumn{2}{c}{$E_{\rm{s-t},n=2}^{\rm exact}$} &  \multicolumn{2}{c}{$E_{\rm{s-t},n=2}^{\rm pert.}$}&  \multicolumn{2}{c}{ $E_{\rm{Total}}^{\rm pert.}$}&\multicolumn{2}{c}{Difference}\\
\hline
2  &  0&.013992 & 0&.013992  & 0&.01521  &0&.00122\\
3  &  0&.019965 & 0&.019965  & 0&.02060  &0&.00065\\
4  &  0&.022656 & 0&.022656  & 0&.02306  &0&.00040\\
5  &  0&.024124 & 0&.024125  & 0&.02439  &0&.00026\\
6  &  0&.025026 & 0&.025028  & 0&.02522  &0&.00019\\
7  &  0&.025604 & 0&.025610  & 0&.02581  &0&.00020\\
8  &  0&.025985 & 0&.025997  & 0&.02624  &0&.00024\\
9  &  0&.026197 & 0&.026222  & 0&.02658  &0&.00035\\
10 &  0&.026290 & 0&.026335  & 0&.02684  &0&.00051\\
\hline
\hline
\end{tabular}
\end{table}

The various contributions in Table \ref{tableI} are as follows: The first three terms give the mass scaling, and the first- and second-order mass polarization corrections due to the $H_{\rm MP}$ term in the nonrelativistic Hamiltonian. These terms cancel exactly from the SIS.  The main contributions to the SIS come from the singlet-triplet mixing term, relativistic recoil (which includes the Stone term), and relativistic mass polarization.  The anomalous magnetic moment terms contribute to the SIS at the 20\,kHz level. The electron-nucleus QED recoil terms terms $E_{\rm M,1} + E_{\rm R,1}$ contribute at the level of 2.7\,MHz, but they are $J$-independent (in lowest order), and so do not contribute to the SIS.  The electron-electron QED recoil terms $E_{\rm L,2}$ are substantially smaller and $J$-independent. The term denoted ``$\alpha^2(\mu/M)^2$ mass scaling" includes quadratic mass scaling of the Breit interaction, and the mass polarization correction to the Stone term. ``$\alpha^2(\mu/M)^2$ second-order M.P." denotes  quadratic mass polarization corrections to the Breit interaction, denoted by $\Delta E_X^{(2)}$ in Eq.\,(\ref{quadratic}).

\begin{table}[b]
\caption{$J$-dependent contributions to the $^{13}$C$^{4+}-^{12}$C$^{4+}$ splitting isotope shift (SIS) in the $1s2p\;^3\!P_J$ fine-structure multiplet relative to $J=0$. Units are MHz.}
\label{tableIII}
\begin{tabular}{l r@{}l r@{}l r@{}l}
\hline
\hline
\rule{0cm}{3mm}Term  &
\multicolumn{2}{c}{$2\;^3\!P_1$}&  \multicolumn{2}{c}{$2\;^3\!P_2$}\\
\hline
$\alpha^2\mu/M$ relativistic recoil       &$      6$&.1686     &$     68$&.5812   \\ 
$\alpha^2\mu/M$ rel.\ mass pol.\          &$     -1$&.3284     &$    -40$&.2567   \\ 
$\alpha^3\mu/M$ anom.\ mag.\ moment        &$      0$&.0014     &$      0$&.0363   \\ 
$\alpha^4\mu/M + \cdots$ S.I.             &$     -0$&.0002     &$     -0$&.0006   \\ 
$\alpha^2(\mu/M)^2$ second-order M.P.     &$      0$&.00035    &$     -0$&.0042   \\ 
$\alpha^2(\mu/M)^2$ mass scaling          &$      0$&.0048     &$      0$&.0099   \\ 
$\alpha^4\mu/M$ $n=2$ sing.-trip.\ mixing &$     -2$&.7676     &$      0$&.0000   \\ 
$\alpha^4\mu/M$ remainder                 &$     -$0&.021(1)   &$     -0$&.001(1) \\ 
Total                                     &$      2$&.058(1)   &$     28$&.365(2) \\ 
\hline
\hline
\end{tabular}
\end{table}

The singlet-triplet mixing term requires further discussion.  As explained in Sec.\,\ref{sec:Theory}, the dominant contribution to the sum over states comes from the $n = 2$ term.  Table \ref{tableII} compares this single contribution with the total of all contributions of order $\alpha^4\mu/M$, as calculated by Drake for $Z=2$ \cite{Drake2002}, and Pachucki \cite{Pachucki2010} for $2\le Z \le 10$. The second column labelled  $E_{\rm{s-t}}^{\rm exact}(n=2)$ is the isotope shift to the $2\,^3\!P_1$ state obtained by exact diagonalization of the complete Breit interaction (including the Stone term) in the two-dimensional basis set formed by the nonrelativistic $2\,^1\!P_1$ and $2\,^3\!P_1$ states. 
The third column labelled $E_{\rm{s-t},n=2}^{\rm pert}$ is the same quantity calculated by perturbation theory from just the $n=2$ term in Eq.\,\eqref{eq:pert}.  The two are in close agreement up to $Z=6$, but then begin to diverge since perturbation theory over-estimates the energy shift relative to exact diagonalization. The fourth column is the total contribution of order $\alpha^4\mu/M$, summed over all $n$ and including the Douglas and Kroll terms; and the last column is the difference $E_{\rm Total}^{\rm pert.} -E_{\rm{s-t},n=2}^{\rm pert}$.  
The difference reaches a minimum of only 0.75\% at $Z=6$ and then starts increasing again.  The strategy is therefore to subtract out the $n=2$ perturbative contribution to $E_{\rm Total}^{\rm pert.}$, and replace it with $E_{\rm{s-t}}^{\rm exact}(n=2)$, as listed as a separate contribution in Table \ref{tableI}. The remainder of 0.00019 $Z^8\alpha^4\mu/M$ a.u.\ is included in the third last row in Table\,\ref{tableI}  as the ``$\alpha^4\mu/M$ remainder" of $-0.021(1)$ MHz. The difference $E_{\rm{s-t}}^{\rm exact}(n=2)-E_{\rm{s-t}}^{\rm pert}(n=2)$ is only 1\,kHz for the $^{12,13}$C SIS, but it increases rapidly with $Z$ to 7\,kHz for $^{16,17}$O, and 100\,kHz for $^{19,20}$Ne.

\begin{table}[t]
\caption{Theoretical values of the SIS in the heliumlike ions for the naturally abundant isotopes of helium ($^{3,4}$He$^{}$) to oxygen ($^{16,18}$O$^{6+}$), and experimental values for helium \cite{Zhao.1991}, lithium \cite{Riis.1994} and carbon (this work). The differences in the splitting intervals $\Delta\nu_{01}$, $\Delta\nu_{02}$ of the isotope pair are relative to the $1s2p\;^3\!P_0$ state.  Units are MHz.}
\label{tab:exp_theory_P0}
\begin{tabular}{c r@{}l r@{}l r@{}l r@{}l}
\hline
\hline
        & \multicolumn{4}{c}{$\Delta\nu_{01}$}&  \multicolumn{4}{c}{$\Delta\nu_{02}$}\\
Isotopes &\multicolumn{2}{c}{Theory} & \multicolumn{2}{c}{Exp.}& \multicolumn{2}{c}{Theory}& \multicolumn{2}{c}{Exp.}\\
\hline
$^{4}$He$-^{3}$He               &$  -0$&.271       &$    -0$&$.28\,(5) $&$         0$&$.721$ &$    0$&$.70\,(5)$   \\
$^{7}$Li$^+-^{6}$Li$^+$         &$  -0$&.292       &$    -0$&$.3\,(9)  $&$        2$&$.558$ &$    1$&$.2\,(8)$\\
$^{10}$Be$^{2+}-^{9}$Be$^{2+}$  &$   0$&.093       &$     $&&$         6$&$.024\,(1)$\\
$^{11}$B$^{3+}-^{10}$B$^{3+}$ &$    1$&$.092     $& & &$     15$&$.951\,(5)$ \\
$^{13}$C$^{4+}-^{12}$C$^{4+}$ &$    2$&.058\,(1)  &$    5$&$.1\,(3.0) $&$    28$&$.37\,(1) $&$    30$&$.3\,(3.1)$  \\
$^{14}$C$^{4+}-^{12}$C$^{4+}$ &$    3$&$.815\,(2)$\rule{1mm}{0mm}&$    6$&$.0\,(4.4)$\rule{1mm}{0mm} &$    52$&.59\,(2) &$    55$&$.1\,(4.2)$ \\
$^{15}$N$^{5+}-^{14}$N$^{5+}$   &$   1$&.671\,(3)  &$     $&&$         44$&$.21\,(3)$\\
$^{18}$O$^{6+}-^{16}$O$^{6+}$   &$   -4$&.26\,(2)  &$     $&&$        121$&$.8\,(1)$\\
\hline
\hline
\end{tabular}\\
\end{table}

Table \ref{tableIII} summarizes the residual $J$-dependent contributions to the SIS, expressed relative to the $1s2p\;^3P_0$ state. The totals are the values of the SIS to be compared with experiment and are quoted in Table \ref{tab:exp_theory_P0} for the fine-structure shifts in the heliumlike systems with $2\leq Z \leq 8$ of the naturally occurring isotopes  with respect to the most abundant one. 
Moreover, Table\,\ref{tableIV} collects together the values of the SIS coefficients for all the He-like ions with $2\le Z \le 10$.  There is severe numerical cancellation for $\Delta\nu_{01}$, as manifested by two sign changes along the isoelectronic sequence. Other cases can be readily calculated from the SIS coefficients listed in Table \ref{tableIV}.

\begin{table}[b]
\caption{SIS coefficients for the $J=1\rightarrow0$ and $J=2\rightarrow0$ fine-structure transitions of the $1s2p\;^3P_J$ states of heliumlike ions, in units of $Z^5$\,MHz$\cdot$u for $\Delta\nu_{0j}^{(1)}$ and $Z^5$\,MHz$\cdot$u$^2$ for $\Delta\nu_{0j}^{(2)}$.\protect
\footnote{For example, for $Z=6$, the $^{13}$C$^{4+}-^{12}$C$^{4+}$ SIS is $\Delta\nu_{01} = \left(\displaystyle\frac{1}{13.00335}-\frac{1}{12}\right)6^5(-0.04106) + \left(\displaystyle\frac{1}{13.00335^2}-\frac{1}{12^2}\right)6^5(-0.00064)= 2.058$\,MHz.}}
\label{tableIV}
\begin{tabular}{r r@{}l r@{}l r@{}lr@{}l}
\hline
\hline
$Z$ &  \multicolumn{2}{c}{ $\Delta\nu_{01}^{(1)}$} & \multicolumn{2}{c}{ $\Delta\nu_{01}^{(2)}$} & 
 \multicolumn{2}{c}{ $\Delta\nu_{02}^{(1)}$} & \multicolumn{2}{c}{ $\Delta\nu_{02}^{(2)}$} \\
\hline
2   &  0&.10560(1) &--0&.00321 & --0&.2729(1)& --0&.00492\\
3   &  0&.05131(1) &--0&.00227 & --0&.4429(1)& --0&.00335\\
4   &--0&.00790(1) &--0&.00144 & --0&.5304(1)& --0&.00199\\
5   &--0&.03847(1) &--0&.00094 & --0&.5645(2)& --0&.00120\\
6   &--0&.04106(2) &--0&.00064 & --0&.5646(2)& --0&.00072\\
7   &--0&.02088(4) &--0&.00045 & --0&.5541(4)& --0&.00043\\
8   &  0&.01876(5) &--0&.00032 & --0&.5338(5)& --0&.00024\\
9   &  0&.07678(7) &--0&.00024 & --0&.5107(7)& --0&.00012\\
10  &  0&.1528(10) &--0&.00017 & --0&.4870(10)&--0&.00003\\
\hline
\hline
\end{tabular}
\end{table}

Referencing the SIS to the $^3\!P_0$ state as done in Table \ref{tab:exp_theory_P0}, has the advantage that the influence of the singlet-triplet mixing is restricted to the $^3\!P_1$ level that mixes with the $^1\!P_1$ level as discussed above. It does, however, not provide the SIS with the highest experimental accuracy. This is rather achieved if the SIS is calculated with respect to the center of gravity of all $^3\!P_J$ states. The corresponding values are listed in  Table \ref{tab:exp_theory_cog} and again compared to theory. 
Comparing the \Cq{14}-\Cq{12} SIS shows excellent agreement.
This provides a test for the theoretical calculation, though the experimental accuracy can by far not compete with that of theory. 
For \Cq{13}-\Cq{12}, however, the situation is different. While the impact on the center-of-gravity is small, hyperfine-induced fine-structure mixing affects all fine-structure states and shifts the three \lines\ fine-structure transitions by as much as 2\,GHz as presented in \cite{Mueller.2025}.

\begin{table}[t]
    \centering
    \caption{$J$-dependent contributions to the $1s2p\,^3\!P_J$ isotope shifts relative to the center-of-gravity frequencies. In \Cq{13} hyperfine-induced fine-structure mixing mixing affects the transition frequencies of all fine-structure lines. This was explicitly included in the hyperfine analysis labeled (Exp.), while for (Exp.*) a normal hyperfine analysis, neglecting this second-order effect, was performed. }
    \begin{ruledtabular}
    \begin{tabular}{l l l l l}
         Isotope pair & & $1s2p\,^3\!P_0$ & $1s2p\,^3\!P_1$ & $1s2p\,^3\!P_2$ \\
       \hline
       \rule{0mm}{4mm}\Cq{13}-\Cq{12} & Theory &$  -16.446\,(2)$&$  -14.388\,(2)$&$ 11.922\,(2)$ \\
                       & Exp.   &$  -18.5\,(2.4)$&$  -13.4\,(1.8)$&$ 11.8\,(1.2)$ \\
                       & Exp.*  &$  1719.4\,(2.5)$&$ -665.1\,(1.8)$&$ 55.2\,(1.2)$ \\
       \Cq{14}-\Cq{12} & Theory &$  -30.489\,(2)$&$  -26.674\,(2)$&$ 22.102\,(2)$ \\
                       & Exp.   &$  -32.6\,(3.4)$&$  -26.6\,(2.4)$&$ 22.5\,(1.5)$ \\
    \end{tabular}
    \end{ruledtabular}
    \label{tab:exp_theory_cog}
\end{table}

In Table\,\ref{tab:exp_theory_cog}, the experimentally determined $J$-dependent contributions to the isotope shift are given with and without including hyperfine-induced mixing. While the ``normal'' hyperfine analysis without consideration of hyperfine-induced mixing yields deviations by up to 1.7 GHz, excellent agreement is found when including hyperfine-induced mixing explicitly by applying the theoretical magnetic dipole matrix elements tabulated in \cite{Johnson.1997}.

In summary, we have measured and calculated the splitting isotope shift (SIS) in heliumlike carbon ions of the isotopes \Cq{13,14} with respect to \Cq{12} and found excellent agreement. This is of special importance for \Cq{13}, where hyperfine-induced fine-structure mixing is present. The agreement provides a clear validation of the applied correction methods and gives additional confidence to the results presented in \cite{Mueller.2025}.
Results provided for other cases of light isotopes will be helpful to test the consistency of experimental results in upcoming measurements. Even though our results surpass all previous measurements on the splitting isotope shift in He-like \Cq{} ions by orders of magnitudes, a further increase of experimental accuracy would be desirable for more accurate tests of the underlying theory.

\begin{acknowledgments}
We acknowledge support by the Deutsche Forschungsgemeinschaft (DFG, German Research Foundation) -- Project-ID 279384907 -- SFB 1245 and by the BMBF under Contract Nos. 05P19RDFN1.  GWFD acknowledges support by the Natural Sciences and Engineering Research Council of Canada (NSERC) and by the Digital Research Alliance of Canada/Compute Ontario.
\end{acknowledgments}


\noindent

\bibliography{Literature.bib}

\end{document}